\documentclass{article}

\usepackage{arxiv}

\usepackage[utf8]{inputenc} % allow utf-8 input
\usepackage[english]{babel}
\usepackage[T1]{fontenc}    % use 8-bit T1 fonts
\usepackage{hyperref}       % hyperlinks
\usepackage{url}            % simple URL typesetting
\usepackage{booktabs}       % professional-quality tables
\usepackage{amsfonts}       % blackboard math symbols
\usepackage{nicefrac}       % compact symbols for 1/2, etc.
\usepackage{microtype}      % microtypography
\usepackage{lipsum}
\usepackage{microtype}      % microtypography
\usepackage{lipsum}
\usepackage{amsmath}
\usepackage{graphicx}
\usepackage{epstopdf}
\usepackage{cite}

\title{Critical coupling vortex with grating-induced\\ high quality optical Tamm states}

\author{Rashid G. Bikbaev\\
Kirensky Institute of Physics, Federal Research Center KSC SB RAS, Krasnoyarsk 660036, Russia\\
Siberian Federal University, Krasnoyarsk 660041, Russia \\
\And
Dmitrii N. Maksimov\\
Kirensky Institute of Physics, Federal Research Center KSC SB RAS, Krasnoyarsk 660036, Russia\\
Siberian Federal University, Krasnoyarsk 660041, Russia \\
\And
Pavel S. Pankin\\
Kirensky Institute of Physics, Federal Research Center KSC SB RAS, Krasnoyarsk 660036, Russia\\
Siberian Federal University, Krasnoyarsk 660041, Russia \\
\And
Kuo-Ping Chen\\
Institute of
Imaging and Biomedical Photonics, National Chiao Tung University, 71150 Tainan, Taiwan\\
\And
Ivan V. Timofeev\\
Kirensky Institute of Physics, Federal Research Center KSC SB RAS, Krasnoyarsk 660036, Russia\\
Siberian Federal University, Krasnoyarsk 660041, Russia \\
}

\begin{document}
\maketitle

\begin{abstract}
We investigate optical Tamm states supported by a dielectric grating
placed on top of a distributed Bragg reflector. It is found that
under certain conditions the Tamm state may become a bound state in the continuum. The bound state, in its turn, induces the effect of critical
coupling with reflectance amplitude reaching an exact zero. We demonstrate that
the critical coupling point is located in the core of a vortex of the
reflection amplitude gradient in space of wavelength and angle of incidence. The emergence of the vortex is explained by the coupled mode theory.

\end{abstract}

% keywords can be removed
%\keywords{Tаммовский плазмон-поляритон, органический солнечный элемент, фоточувствительный слой, локализация света}

%\section{Introduction}

The combination of plasmon and photon structures leads to  hybrid systems, which have been of great interest  over the recent few years.
One example of such a hybrid system is a two-dimensional lattice of nanoparticles combined with a Fabry-Perot resonator~\cite{Ameling2013,Gerasimov2019EngineeringCrystal}. 
The unique properties of such systems can be applied to narrowband absorbers~\cite{Chanda2011, Liu2015d}, lasers~\cite{zhou2013a,Mischok2018}, sensors \cite{Bahramipanah2015, Chen2018} and photodetectors~\cite{Thompson2017}. The hybrid systems are used to reduce losses in plasmonics ~\cite{Kamakura2017} and for local amplification of the electromagnetic field~\cite{Alrasheed2017}.

Hybrid two-dimensional lattices not only act as structural elements but can also be used as mirrors for engineering localized states. It is recently shown \cite{Wang2018} that a Tamm plasmon (TP)~\cite{Kaliteevski2007} can be excited at the interface between binary Au nanodisk arrays and distributed Bragg reflector (DBR).
In this case coupling between the TP and localized lattice resonance mode results in  dual Tamm states. 
The described structure can be used to measure the Zak phase.
Thus in \cite{Wang2016} the Zak phase has been determined through the interface state. 
% Besides that, the authors showed the manipulation of interface states including both the excitation frequency and the polarization by introducing metasurfaces. {\bf I do not think we need this sentence. }
The propagation of the Tamm plasmons was investigated in  \cite{Chestnov2017} by depositing metallic stripes on top of a semiconductor Bragg mirror. 
It was shown that the TPs are coupled to surface plasmons arising at the stripe edges. These plasmons form an interference pattern close to the bottom surface of the stripe resulting in a change of both the energy and loss rate for the TP. 
\begin{figure*}    \centering
    \includegraphics[width=0.7\textwidth]{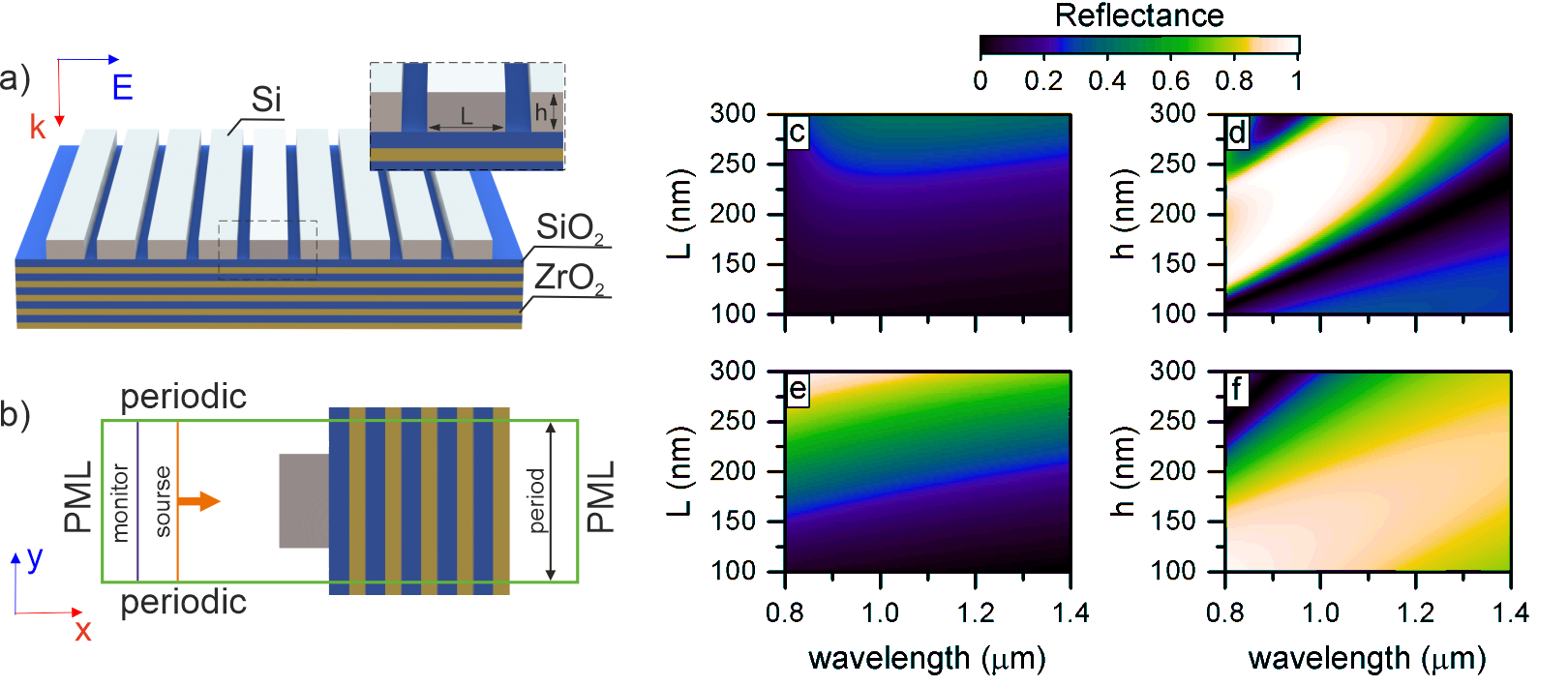}
    \caption{(a) Schematic representation of the structure.(b) Sketch view of the simulation box. Reflectance spectra of the 2D silicon (c,d) and gold (e,f) stripes array for different value of thickness $L$ (c,e) ($h=100$~nm) and $h$ (d,f) ($L=300$~nm). Period of the array $p=400$~nm. }
    \label{fig:fig1}
\end{figure*}
The disadvantage of the hybrid systems is material losses. 
Therefore, dielectric metasurfaces are increasingly used.
For example in \cite{Lee2019} the authors propose narrowband perfect absorbers with enormously high fabrication tolerance, which consist of a low-contrast grating and a finite DBR layer with an ultrathin absorbing medium (graphene).
%It should be noted that the dielectric metasurfaces can be used not only in the visible region of the spectrum but also in the terahertz region. 
%Thus in \cite{Mitrofanov2018} the authors showed that the use of such a metasurface can increase the absorption in the photodetector.
A planar array of optical band-pass filters composed of low loss dielectric metasurface layers sandwiched between two DBRs is reported in \cite{Horie2016}. 

In this paper we demonstrate the advantage of Bragg reflection over total internal reflection.
We show that optical Tamm states (OTS) have better quality factor in comparison with TP and surface plasmon resonance. 
This is because of the absence of material losses and larger resonant volume due to Bragg reflectors \cite {Symonds2017}. Ultimately, in dielectric structure the OTS can obtain
infinite quality factor, i.e. the OTS becomes optical bound state
in the continuum (BIC) \cite{Hsu16}. Such states supported by dielectric
gratings have been previously reported both experimentally \cite{Sadrieva17,Doeleman2018} and theoretically \cite{Bulgakov18, Bulgakov18a}. Here we demonstrate that the Tamm BIC induces the effect of critical coupling (CC).  We show that
the CC corresponds a phase singularity in the space of incidence angle and frequency, while the phase gradient forms a vortex with the CC point in the core. The phase singularities carry a topological charge \cite{Bliokh19} defined as the number of $2\pi$ windings of the phase along a closed contour around the CC point. Based on the temporal coupled mode theory \cite{Fan03} we provide semi-analytic expression for the reflection amplitudes that matches with full-wave numerical simulations. 
The CC points are known to lead to very fast  local variation of the phase of the reflection amplitude \cite{Tsurimaki18}. Here we find that such fast
variation in the vicinity of the CC is unavoidable due to the phase singularity. 
%Also we suggest that the small tangential propagation speed of this state in comparison  with Bloch surface waves~\cite{Abrashitova2018} can be used for creation of delay lines. 
%This advantage can be used in topologically protected transport \cite{Guo2019} of interfacial states.

Let us consider a DBR with a 2D array of silicon nanostripes on top (Fig.~\ref{fig:fig1}a). 
The DBR unit cell is formed of two layers: silica dioxide (SiO$_2$) with thickness $d_a=165$~nm and permittivity $\varepsilon_a=2.1$, and zirconium dioxide (ZrO$_2$) with thickness $d_b=135$~nm and permittivity $\varepsilon_b=4.16$. 
The 2D structure with thickness $h$ and width $L$ has infinite length along $y$ axis with period $p$.
The direction of the incident light on the structure and its polarization are shown by the red and blue arrows, respectively.
The optical properties of the structures have been calculated by commercial Finite-Difference Time-Domain (FDTD) package.
Standard numerical protocols are implemented to mimic infinite 2D periodic structures.
The simulation box is shown in figure~\ref{fig:fig1}b.  
%PhC structure are illuminated from the top by the plane wave with normal incidence along $z$ axis and polarization along $x$ axis. 
%Reflectance $R$ has been calculated at the top of the simulation box. 
%Periodic boundary conditions have been applied at the lateral boundaries of the simulation box, while perfectly matched layer (PML) boundary conditions were used on the remaining top and bottom sides (see. Fig~\ref{fig:fig1}b). 
%An adaptive mesh has been used to reproduce accurately the nanostripes shape. 

%\section{Results}
%\subsection{Reflectance from bare 2D structure}
A localized state, such as OTS, can be formed at the boundary of reflecting media. 
In our structure, reflection from a DBR is provided by the photonic band gap. %, the position of which depends on the thickness and dielectric permittivity of its layers {\bf Could we remove the clause?}. 
If the number of layers is infinite
the layered DBR forms an ideal Bragg reflector \cite{WeiHsu2013} which
prohibits radiation losses.
Reflection from the 2D structure can be controlled by varying the lattice period, as well as the thickness and width of the nanostripes. 
First we calculated the reflectance spectra of the bare 2D structure without a DBR. 
The simulation results are shown in figure~\ref{fig:fig1}(c-f).

\begin{figure}[h]
    \centering
    \includegraphics[scale = 0.65]{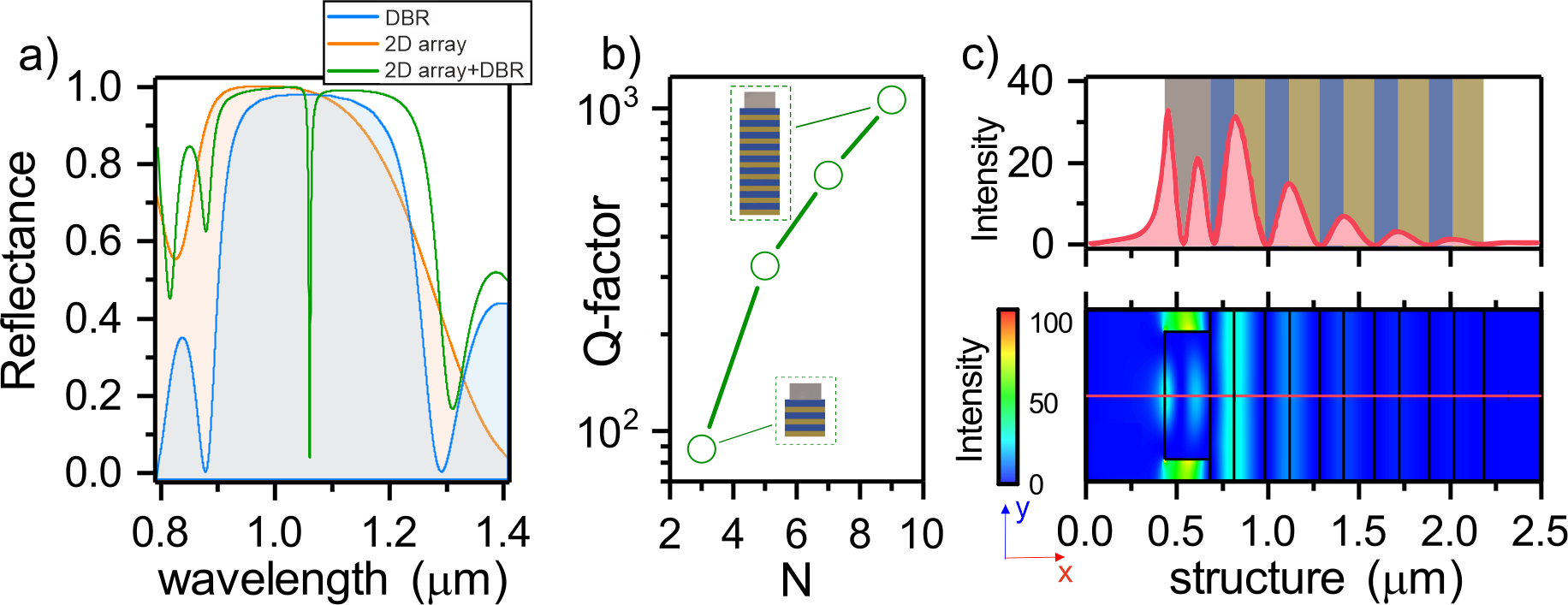}
    \caption{(a) Reflectance spectra of the structure. (b) The Q-factor of the OTS depends on the number of DBR layers and (c) field distribution at the OTS wavelength. }
    \label{fig:fig3}
\end{figure}

The calculations were carried out in two steps. First, we fixed the height of the stripes and varied their width. Thus, the width at which the most reflection from the structure is achieved was determined. In the next step, we varied the height of the stripe with fixed width. In this way, we were able to determine the parameters of the structure at which the largest reflection from the structure can be achieved.
The calculation results showed that the largest reflection is achieved in the range from 800 to 1200 nm for a structure with a stripes thickness $h=250$~nm and a width of $L=300$~nm (see Fig.~\ref{fig:fig1}(c,d)).
We also calculated the reflectance spectra of a 2D array of gold stripes. The results are presented in the figure~\ref{fig:fig1}(e,f), which shows that the structure reflects in the entire range of wavelengths investigated, but the reflectance is not high, and reaches only 60-70\%. We determined that the reflection reaches the maximum value for stripes with thickness $h=150$~nm and width $L=300$~nm.

%\subsection{2D+PhC}
We expect that conjugation of the periodic structures with a DBR could lead to the formation of localized states. To verify this the reflectance spectra of the composite structures were calculated~(see Fig.~\ref{fig:fig3}). Figure \ref{fig:fig3}a shows that DBR band gap is in the range of wavelengths from 900 to 1100~nm overlaps with the reflection region of a 2D silicon stripes structure. 
In the area of their overlap, a narrow spectral line ($\lambda=1060$~nm) is observed within the photonic band gap. 
This line corresponds to a state localised at the interface between DBR and 2D array of silicon stripes.

\begin{figure*}
    \centering
    \includegraphics[width=0.8\textwidth]{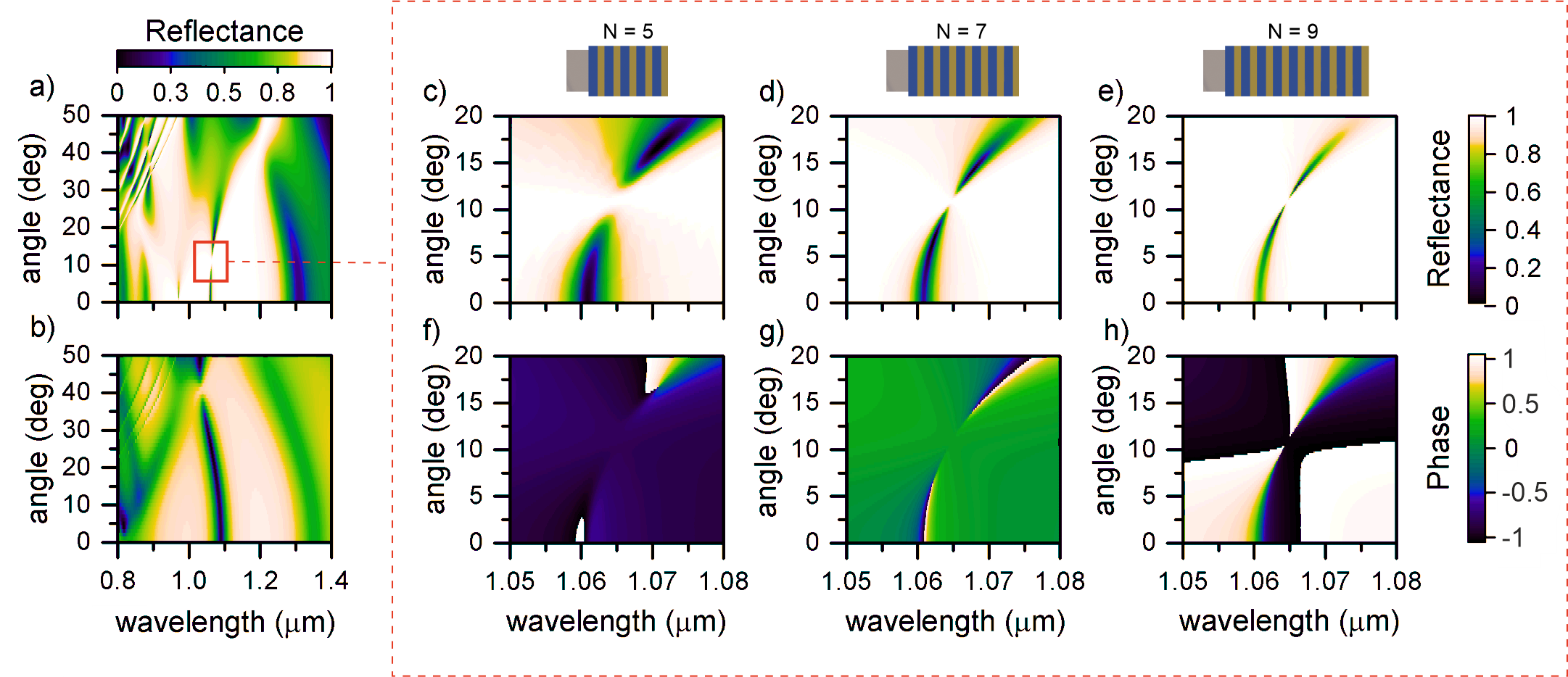}
    \caption{Reflectance spectra of the structure in case of (a) silicon and (b) gold stripes. (c,d,e) Reflectance spectra and (f,g,h) reflection phase  of the DBR bounded by 2D silicon stripes for different number of DBR periods.}
    \label{fig:fig3new}
\end{figure*}
It is important to note that an increase in the number of periods of the DBR leads to a significant increase in the Q-factor of the resonance (see Fig.~\ref{fig:fig3}b). Thus, when increasing $N$ from 3 to 9, the Q-factor grows by more than an order of magnitude.
This because the increase in the number of periods of the DBR leads to a decrease of the energy relaxation rate in the lower half-space (the transmission channel of the DBR). When the DBR is infinite, the losses
into the lower half-space are totally suppressed by the photonic band gap, whereas the losses to the upper half-space are suppressed by interference \cite{WeiHsu2013}. Thus, the OTS becomes a BIC at a specific incidence angle $\theta \approx 10.9^{\circ}$.  
\textbf{}
%According to the temporal coupled mode theory, the Q-factor of the line %is due to the leakage of energy into the channels. As a result, %reducing energy leakage to one of the channels increases the Q-factor %of the resonant line.

The distribution of the field intensity at the wavelength of the localized state is shown in the figures~\ref{fig:fig3}c .
On can see that the field is localized close to the interface 
between the two constituent subsystems and exponentially decreases in the photonic crystal and the 2D lattice.
However, unlike conventional OTS, in our case the field is also localized on the front border of the 2D medium.
%\textcolor{red}{Explanation}
%The distribution of the field in the plane perpendicular to the axis of the strips is shown in the figure~\ref{fig:fig3}c. 
It is interesting to note that the maximum intensity of the field is achieved not in the center of the stripe, but at its boundaries (see bottom subplot in Fig.~\ref{fig:fig3}c). 
This leads to enhancement of the field in the inter-stripe space.

%The comparative reflection spectra of the structure for PhC bounded by with a 2D lattice of gold or silicon strips are shown in figure~\ref{fig:fig4}.

%The figure shows that the use of lossless material leads to the formation of a higher-quality state. This makes it possible to assume that the propagation length in the case of an oblique incidence of light, will be greater for the states of localised at the interface between PhC and silicon strips, since the propagation length of the surface wave is proportional to its Q-factor. 

%\begin{figure}
%\begin{minipage}[b]{0.24\textwidth}
%\includegraphics[scale = 0.6]{Fig5a.png}\\
%\end{minipage}
%\hfill
%\begin{minipage}[b]{0.25\textwidth}
%\includegraphics[scale = 0.6]{Fig5b.png}\\
%\end{minipage}
%\vfill
%\begin{minipage}[b]{0.24\textwidth}
%\includegraphics[scale = 0.6]{Fig5c.png}\\
%\end{minipage}
%\hfill
%\begin{minipage}[b]{0.25\textwidth}
%\includegraphics[scale = 0.6]{Fig5d.png}\\
%\end{minipage}
      % \caption{(a,c) Reflectance spectrum and reflection phase calculated with FDTD. (b,d) Reflectance spectrum and reflection phase calculated with TCMT (Eq.\ref{reflection}).}
  %  \label{fig:figTCMT}
%\end{figure}
The angular dependencies of the reflectance spectra are shown in the figure~\ref{fig:fig3new}.
The figure shows that the spectra of the localized state for DBR bounded by 2D silicon or gold structure are cardinally different (see Fig.~\ref{fig:fig3new}(a,b)).
The difference is not only in the width of the resonance, but also in the shifting inside the photonic band gap. So, for gold stripes, the localized state is shifted to the blue, same as the conventional TP, while for silicon stripes structure, we observe a red shift. Moreover, in both cases, we see the collapse of the resonant lines, with the only difference that for silicon stripe structure, this effect is observed at lower angles.
Let's look at the dielectric structure in more detail.
We calculated the reflection spectra of the structure and phase of the reflected wave in the region of the collapse of the resonance line. The calculation results are shown in figures \ref{fig:fig3new}(c-h).

 It can be seen from the reflection spectra that an increase of $N$ leads to the contraction of the resonant lines in the BIC point. It is important to note that for $N=5$ (see Fig.~\ref{fig:fig3new}f)  only one critical coupling point is observed, while for $N=7$ and  $N=9$ (see Fig.~\ref{fig:fig3new}(g,h))  two critical coupling points are present. As the calculations have shown, even at $N=9$, we do not see merging of the two CC points. Further increase of $N$ up to infinity will allow us to achieve this effect, but in this case we would have the problem both with simulations and  experimental implementation.
Nevertheless, we see that the variation of $N$ in the specified interval allowed us to demonstrate this effect.

%{\it Coupled mode theory}. 
The scattering spectrum in the vicinity of a localized state can be described in the framework of the temporal coupled mode theory \cite{Fan03}.
According to the temporal
coupled mode theory the reflection amplitude can be written
as follows
\begin{equation}\label{reflection}
    r=e^{i2\eta}\left(-1+\frac{2\gamma_1(\theta)}{i[\omega-
    \bar{\omega}(\theta)]+\gamma_1(\theta)+\gamma_2} \right),
\end{equation}
where $\omega$ is the frequency of the incident wave, $\bar{\omega}$ - the resonant frequency, $\eta$ - global phase, $\theta$ - the angle of incidence, and $\gamma_{1,2}$ are the inverse life-times due to coupling to the upper and
lower half-spaces, correspondingly. Notice, that we ignored the dependence
of $\gamma_2$ on $\theta$ as the coupling to the lower half-space is equally
suppressed by the DBR stop band at all angles of incidence. In the upper half-space though, the coupling is cancelled by destructive interference only if $\theta=\theta_{\rm BIC}$. By setting $r=0$ one immediately arrives at the condition
for the CC
\begin{align}\label{cc}
& \omega=\bar{\omega}(\theta_{\rm CC}), \nonumber \\
& \gamma_2=\gamma_1(\theta_{\rm CC}).
\end{align}
Since we have two parameters to satisfy the two equations above the
CCs are points in the plane of $\theta$ and $\omega$. One can immidiately see from Eq. (\ref{cc}) that in case of a true BIC, $N=\infty$ and $\gamma_2=0$ the two CC points merge with the BIC. 
The dispersion of the entries of Eq. (\ref{reflection})
are given by \cite{Bulgakov17}
\begin{align}\label{Taylor}
    & \bar{\omega}=\omega_0-\alpha \theta^2 +{\cal O}(\theta^4), \nonumber \\
    & \gamma_1=\beta(\theta^2-\theta^2_{\rm BIC})^2 +{\cal O}(\theta^6),
\end{align}
where we took into account the symmetry of the leaky band about the $\Gamma$-point. Equation (\ref{Taylor}) contains two unknowns, $\alpha$ and $\beta$. The first, $\alpha$, can be easily found by fitting to the Lorentzian center-frequency in Fig.~\ref{fig:fig3new}d. The second parameter, $\beta$, can be derived as
\begin{equation}\label{beta}
    \beta=\frac{\gamma_2}{(\theta^2_{\rm CC}-\theta^2_{\rm BIC})^2},
\end{equation}
$2\gamma_2$ being the half-width of the Lorentzian at the CC angle. The phase singularity manifests itself in the vortical behaviour of the gradient
of the phase, $\phi$ \cite{Bliokh19} of the reflection coefficient in space of $\theta$ and $\omega$. The phase $\phi$ in Eq. (\ref{gradient}) is implicitly defined via $r=|r|e^{i\phi}$. Although the phase is not uniquely defined due
to $2\pi$ windings around the CC point, the gradient is a well-behaved function
and, thus, can be derived analytically from Eq. (\ref{reflection}), by using the following equation 
\begin{equation}\label{gradient}
    \nabla \phi=\frac{r^*\nabla r - r \nabla r^*}{2i|r|^2}.
\end{equation}
We spare the reader of the cumbersome equations resulting from
substitution of Eq. (\ref{reflection}) into Eq. (\ref{gradient}). Notice, though, that since the global phase $\eta$ is independent of both $\theta$
and $\lambda$ it is absent from Eq. (\ref{gradient}). 

\begin{figure}
    \centering
    \includegraphics[scale = 0.5]{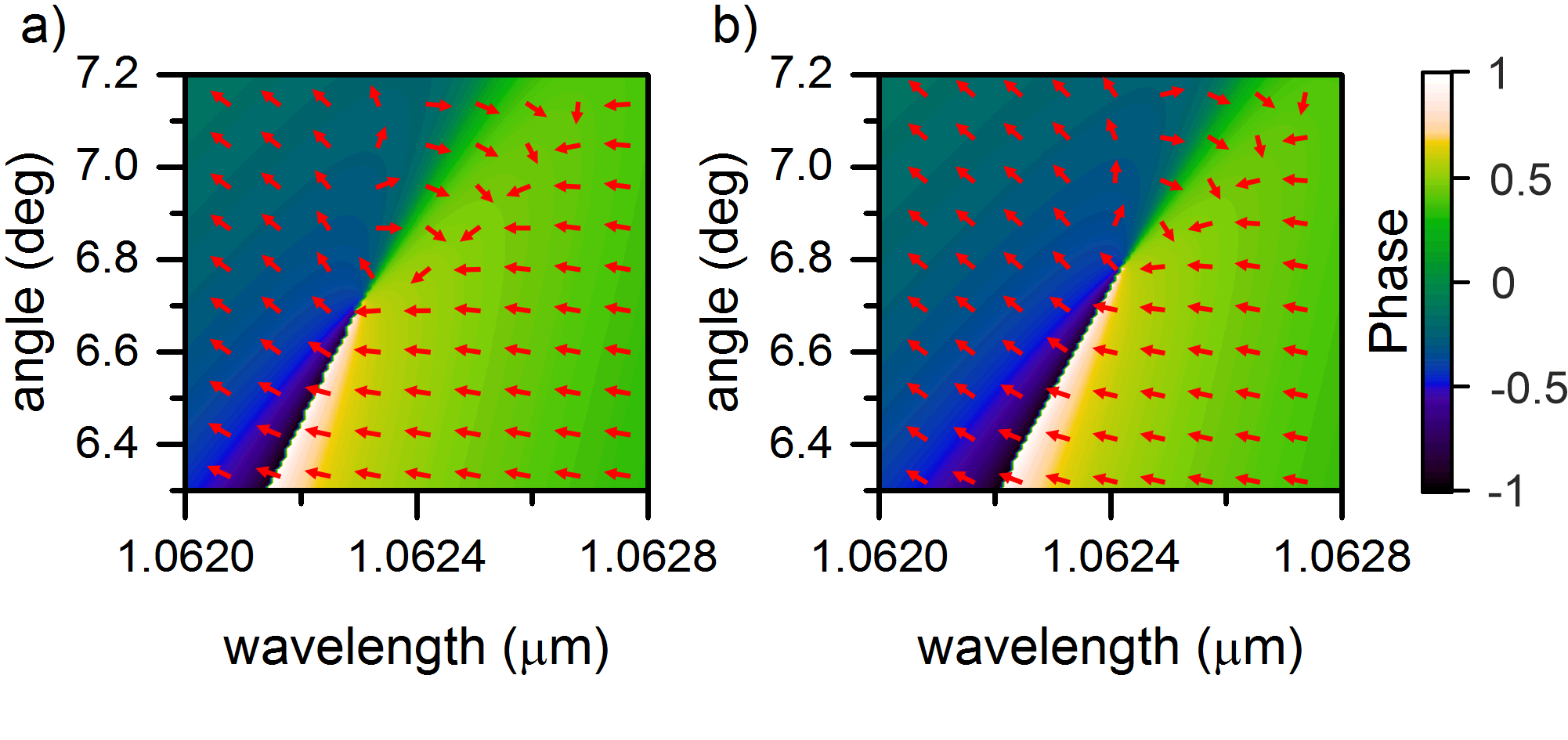}
    \caption{Reflection phase of the DBR bounded by 2D lattice for N=7 obtained by (a) FDTD and (b) coupled mode theory. The red arrows represent the direction of the phase gradient.} 
    \label{fig:fig4new}
\end{figure}
In Fig.~\ref{fig:fig4new} we compare the results obtained with Eq.~(\ref{gradient}) against the results of full-wave numerical simulations. The figure shows the phase of the wave reflected from the structure and the direction of the phase gradient near one of the critical coupling points.
It should be noted that the results are in good agreement and critical coupling vortices are observed in both cases.
%It is important to note that the vortices are not symmetrical and are stretched to the BIC point. {\bf I do not think we need the last sentence.}

In summary, we have demonstrated that the critical coupling effect induced by a Tamm BIC is associated with a phase singularity of the reflection coefficient in the space of wavelength and angle of incidence of the impinging plane wave. The scattering spectrum is explained by a single resonance coupled mode approach. The resulting equation is found to correctly account of the topological property of the critical coupling vortex.  We speculate that the results presented open novel opportunities in engineering critical coupling, whereas the phase singularity yields
unlimited sensitivity to variation of the system's parameters paving a way to sensing applications.

\section*{Acknowledgment}
This work was co-funded by RFBR, project no. 19-52-52006, and the Taiwan Ministry of Science and Technology, project no. 108-2923E-009-003-MY3.

\bibliographystyle{unsrt}
\bibliography{biblio,references, BSC_light_trapping}

% Full bibliography will be added automatically on a new page for Optics Letters submissions. This command is ignored for journal article submissions.
% Note that this extra page will not count against page length.
%\bibliographyfullrefs{biblio,references, BSC_light_trapping}

\end{document}